\documentclass[preprintnumbers,amsmath,11pt,amssymb,floatfix,
superscriptaddress,nofootinbib]{revtex4}
\usepackage{graphicx}
\usepackage{epsfig}
\usepackage{bm}
\usepackage{amsfonts}

\input epsf

\begin{document}

\title{\Large Study of Tachyonic Field and its statefinder diagnostics in Various Scenarios of The Anisotropic Universe}

\author{{\bf Chayan Ranjit}}
\email{chayanranjit@gmail.com} \affiliation{Department of
Mathematics, Seacom Engineering College, Howrah - 711 302, India.}

\author{\bf Surajit Chattopadhyay}
\email{surajit_2008@yahoo.co.in,
surajit.chattopadhyay@rediffmail.com} \affiliation{Department of
Computer Application (Mathematics Section), Pailan College of
Management and Technology, Bengal Pailan Park, Kolkata-700 104,
India.}

\date{\today}

\begin{abstract}
In the present work, we have considered N-dimensional Einstein
field equations in which 4-dimensional space-time is described by
a FRW metric and that of the extra d-dimensions by an Euclidean
metric. Considering the universe filled with tachyonic field we
have reconstructed the potential $V(\phi)$ corresponding to the
field reconstructions in an anisotropic universe under
\emph{Emergent-Power law}, \emph{Emergent-Intermediate},
\emph{Emergent-Logamediate }and \emph{Logamediate-Intermediate}
scenarios. The statefinder parameters have been investigated in
all of the said scenarios.
\end{abstract}

\pacs{}

\maketitle

\section{\bf{Introduction}}
The two independent observational signals on distant Type Ia
supernovae (SNIa) in 1998 have revealed the speeding up of our
universe. This acceleration implies that if the theory of
Einstein's gravity is reliable on cosmological scales, then our
universe is dominated by a mysterious form of matter. This unknown
component possesses some remarkable features, for instance it is
not clustered on large length scales and its pressure must be
negative in order to be able to drive the current acceleration of
the universe \cite{saridakis}. This matter content is dubbed as
``dark energy" (DE) that occupies about $70\%$ of today's
universe. Reviews on DE include \cite{saridakis}, \cite{sami},
\cite{padmanabhan} and \cite{Li}. A chronological history of the
research on DE is discussed in \cite{Li}. The basic characteristic
of DE is that its equation of state (EOS) parameter $w =p/\rho$,
where $\rho$ is the energy density and $p$ is the pressure that
has a negative value. The simplest candidate of dark energy is a
tiny positive cosmological constant \cite{sami} corresponds to a
fluid with a constant equation of state $w=-1$. However, as is
well known, it is plagued by the so-called ``cosmological constant
problem" and ``coincidence problem" \cite{sami}. Other dark energy
models include quintessence \cite{Ratra}, phantom \cite{Najiri},
quintom \cite{saridakis}, Chaplygin gas \cite{Gorini}, tachyon
\cite{chimento}, hessence \cite{Zhao} etc.
\\
The weakness of the gravitational force has been successfully
explained by postulating the existence of extra dimensions
\cite{Arkani-Hamed1}. Today, there is a large variety of promising
models and theories which suggest the existence of more than three
spatial dimensions. Most notably, string theory suggests the
existence of seven additional spatial dimensions
\cite{Poppenhaeger}. As our space-time is explicitly four
dimensional in nature so the `hidden' dimensions must be related
to the dark matter and dark energy which are also 'invisible' in
nature. Model of higher dimension was proposed by Kaluza and Klein
\cite{T. Kaluza,O. Klein} who tried to unify gravity with
electromagnetic interaction by introducing an extra dimension
which is basically an extension of Einstein general relativity in
5D. The activities of extra dimensions also verified from the STM
theory \cite{Randall} proposed recently by Wesson et al.
\cite{Wesson}. The importance of extra dimension in cosmology has
been discussed by many authors. In some analysis the metric is
assumed to be uniform in extra dimension \cite{Arkani-Hamed,Lyth}.
For the metric having non-trivial dependence the extra dimension,
the analysis on inflation may be altered \cite{Lukas,Chamblin}.
Panigrahi and Chatterjee \cite{Panigrahi2008} have found that the
inflationary scenario is possible for inhomogeneous extra
dimensional model and for the homogeneous case an initially
decelerating universe starts accelerating undergoing a flip.
Ib$\acute{a}$nez and Verdaguer \cite{Vardaguer} have obtained a
set of solutions of Einstein's equation in an N-dimensional vacuum
model and also homogeneous solutions with expanding 3-dimensional
isotropic space. The 4D perfect fluid solutions in flat universe
model have been obtained by Krori et al. \cite{Krori}, Gleiser and
Diaz \cite{Gleiser}, using a higher dimensional anisotropic
cosmology (Bianchi-I) which are compatible with contraction of all
the extra dimensions. In \cite{Paul} Paul has considered the
theories of imperfect fluid such as Eckart, EIT, TIS and FIS to
obtain the cosmological solutions for flat FRW with extra
dimensions by Kasner type Euclidean metric and present an analysis
of a n-dimensional vacuum Einstein's field equations. In
\cite{Gorbunov} Gorbunov and Sibiryanov have proposed a
cosmological model of self accelerated brane universe with warped
extra dimension. In  Peng et al. \cite{Peng} extended the direct
quantum approach to the FRW cosmology from 4D to 5D and obtained a
Hamiltonian formulation for a wave like 5D FRW cosmology. In
 Panigrahi et al. \cite{Panigrahi2006} have shown a scenario
in-homogeneous 5D space time which behave a decelerating expansion
in the early epoch along with an accelerating situation at the
present line without introducing any external quintessence-like
scalar field in the presence of extra dimension.
\\
In the present work, we have considered N-dimensional Einstein
field equations in which 4-dimensional space-time is described by
a FRW metric and that of the extra d-dimensions by an Euclidean
metric. Considering the universe filled with tachyonic field we
have reconstructed the potential $V(\phi)$ corresponding to the
field reconstructions in an anisotropic universe. A rolling
tachyon has an interesting equation of state whose parameter
smoothly interpolates between $-1$ and 0 \cite{Gibbons}. This has
led to several attempts to construct viable cosmological models
using the tachyon as a suitable candidate for the inflaton at high
energy \cite{sami}, \cite{Feinstein}, \cite{Qureshi},
\cite{Sami1}, \cite{setare1}, \cite{setare2}. In a flat FRW
background the energy density and pressure of tachyon are given by
\cite{sami}
\begin{equation}
\rho=\frac{V}{\sqrt{1-\dot{\phi}^{2}}}~;~~~~~~p=-V\sqrt{1-\dot{\phi}^{2}}
\end{equation}
where $V$ and $\phi$ denote the potential and tachyonic field
respectively. The tachyonic matter might provide an explanation
for inflation at the early epochs and could contribute to some new
form of cosmological dark matter at late times \cite{Qureshi}. In
isotropic FRW universe, the tachyonic field was studied by
\cite{surajit}, where it was observed by studying the statefinder
diagnostics \cite{sahni} that the tachyonic interpolates between
dust and $\Lambda$CDM stages of the universe. In the present paper
we are going to consider the scale factors $a(t)$ and $b(t)$ in
the following forms:

\begin{enumerate}
    \item $a(t)=A\left(\beta+e^{\alpha t}\right)^{m}$
    with $A>0,~~\alpha>0,~~\beta>0,~~m>1$ (emergent expansion)\cite{Mukherjee,Debnath3}.
    \item $a(t)=exp(A(\ln t)^{m})$ with $Am>0,~~m>1$ (logamediate expansion)\cite{Barrow,Debnath2}.
    \item $b(t)=exp (B t^{n})$ with
    $B>0;~~0<n<1$ (intermediate expansion)\cite{Barrow,Debnath2}.
    \item $b(t)=exp(B(\ln t)^{n})$ with $Bn>0,~~n>1$ (logamediate expansion)\cite{Barrow,Debnath2}.
    \item $b(t)\propto t^{n}$ with $n>1$ (power law expansion)\cite{sami}.
\end{enumerate}

Some authors first choose the scale factor in power law,
exponential or in other forms and then find out other variables
with some conditions under these solutions. This `reverse' way of
investigations had earlier been used extensively by \cite{Ellis}
who chose various forms of scale factor and then found out the
other variables from the field equations. Subsequently, this
approach has been adopted by \cite{Banerjee}, who clearly stated
``This is not the ideal way to find out the dynamics of the
universe, as here the dynamics is assumed and then the fields are
found out without any reference to the origin of the field. But in
the absence of more rigorous ways, this kind of investigations
collectively might finally indicate towards the path where one
really has to search". In another study, reference
\cite{Feinstein} assumed scale factor in the power law form to
model the potential by an inverse square law in terms of the
tachyon field.
\\
In the present work we have adopted this ``reverse approach" to
reconstruct the potential of the tachyonic field in the
anisotropic universe. Subsequently, we have investigated how the
statefinder parameters $\{r,s\}$ behave in this case. We have
chosen combinations of choices of scale factors and investigated
different scenarios under the assumption that the universe is
filled with tachyonic field. The basic equations are discussed in
the following section.
\\\\

\section{\bf{Basic Equations}}
In the present work we consider homogeneous and anisotropic
$N$-dimensional space-time model described by the line
element\cite{Debnath1}

\begin{equation}
ds^{2}=ds^{2}_{FRW}+\sum_{i=1}^{d}b^{2}(t)dx_{i}^{2}
\end{equation}

where $d$ is the number of extra dimensions $(d=N-4)$ and
$ds^{2}_{FRW}$ represents the line element of the FRW metric in
four dimensions is given by

\begin{equation}
ds^{2}_{F R
W}=-dt^{2}+a^{2}(t)\left[\frac{dr^{2}}{1-kr^{2}}+r^{2}(d\theta^{2}+\sin^{2}\theta
d\phi^{2})\right]
\end{equation}

where $a(t)$ and $b(t)$ are the functions of $t$ alone represent
the scale factors of 4-dimensional space time and extra
$d$-dimensions respectively. Here  $k ~(=0, ~\pm 1)$  is  the
curvature  index of the corresponding 3-space, so  that  the above
Universe is
described  as  flat, closed  and  open respectively.\\

The Einstein's field equations for the above non-vacuum higher
dimensional space-time symmetry are

\begin{equation}
3\left(\frac{\dot{a}^{2}+k}{a^{2}}\right)=\frac{\ddot{D}}{D}-\frac{d^{2}}{8}
\frac{\dot{b}^{2}}{b^{2}}+\frac{d}{8}
\frac{\dot{b}^{2}}{b^{2}}+\rho
\end{equation}

\begin{equation}
2\frac{\ddot{a}}{a}+\frac{\dot{a}^{2}+k}{a^{2}}=\frac{\dot{a}}{a}\frac{\dot{D}}{D}+\frac{d^{2}}{8}
\frac{\dot{b}^{2}}{b^{2}}-\frac{d}{8} \frac{\dot{b}^{2}}{b^{2}}-p
\end{equation}
and
\begin{equation}
\frac{\ddot{b}}{b}+3\frac{\dot{a}}{a}\frac{\dot{b}}{b}+\frac{\dot{D}}{D}\frac{\dot{b}}{b}-\frac{\dot{b}^{2}}{b^{2}}=-\frac{p}{2}
\end{equation}

where $\rho$ and $p$ are energy density and isotropic pressure
respectively. Here we choose here $8 \pi G=c=1$ and
$D^{2}=b^{d}(t)$, so we have
$\frac{\dot{D}}{D}=\frac{d}{2}\frac{\dot{b}}{b}$ and
$\frac{\ddot{D}}{D}=\frac{d}{2}\frac{\ddot{b}}{b}+\frac{d^{2}-2d}{4}\frac{\dot{b}^{2}}{b^{2}}$.
\\

Sahni et al \cite{sahni}  proposed the trajectories in the
$\{$r,s$\}$ plane corresponding to different cosmological models
to depict qualitatively different behavior. Since this parameters
are dimensionless so they allow us to characterize the properties
of dark energy in a model independently. The statefinder
diagnostic along with future SNAP observations may perhaps be used
to discriminate between different dark energy models. The above
statefinder parameters are given by \cite{sahni}:

\begin{equation}
r=1+3\frac{\dot{H}}{H^{2}}+\frac{\ddot{H}}{H^{3}} ~~\text{and} ~~
s=\frac{r-1}{3(q-\frac{1}{2})}
\end{equation}
where $q$ is the deceleration parameter defined by
$q=-1-\frac{\dot{H}}{H^{2}}$. In the anisotropic universe under
consideration, the Hubble parameter $H$ is given by
$H=\frac{1}{d+3}(3\frac{\dot{a}}{a}+d\frac{\dot{b}}{b})$~. In this
paper we have reconstructed the statefinder parameters for
different scenarios to be discussed in the subsequent sections.
Finally we graphically analyzed $\{r, s\}$ trajectory for various scenarios.\\

\begin{figure}
\includegraphics[scale=0.5]{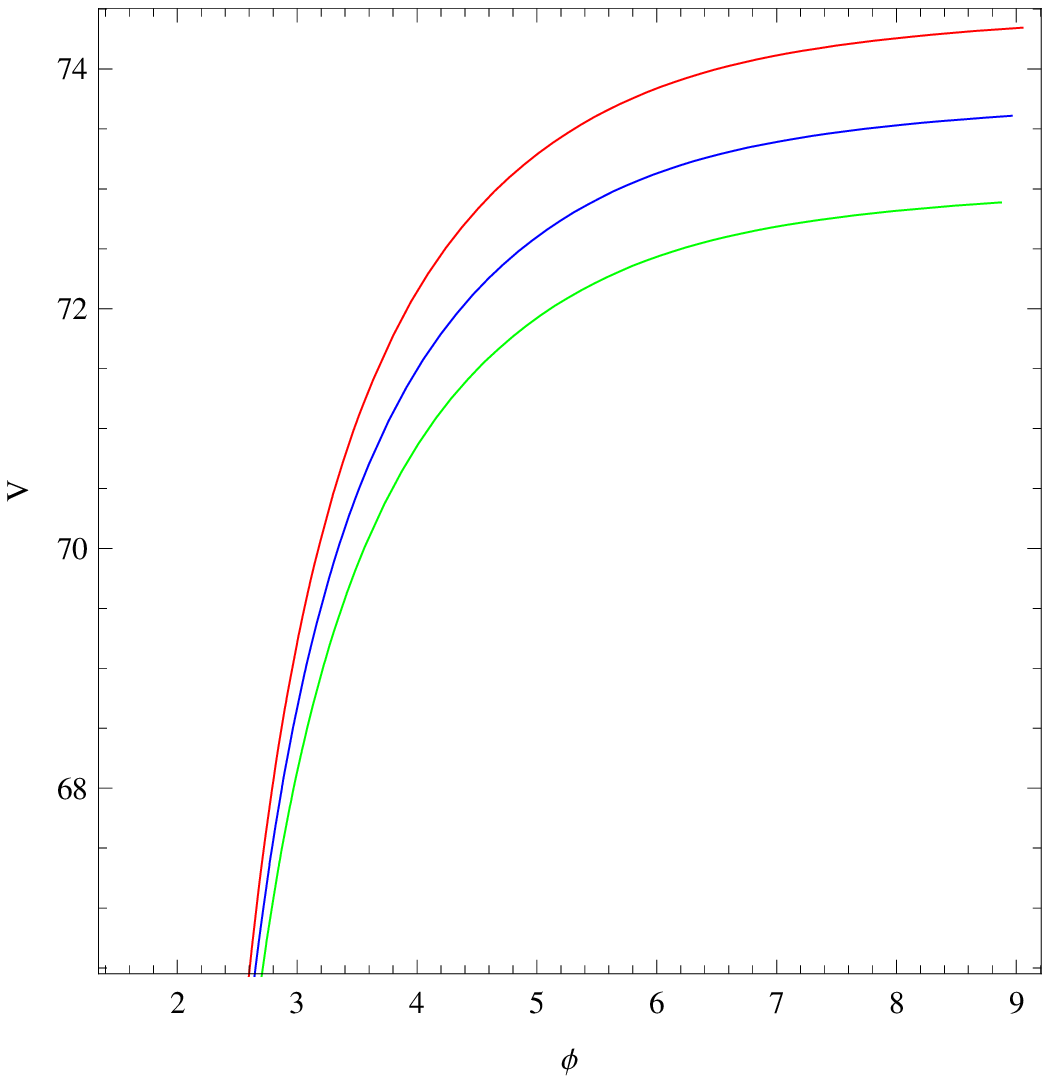}~~~~~
\includegraphics[scale=0.9]{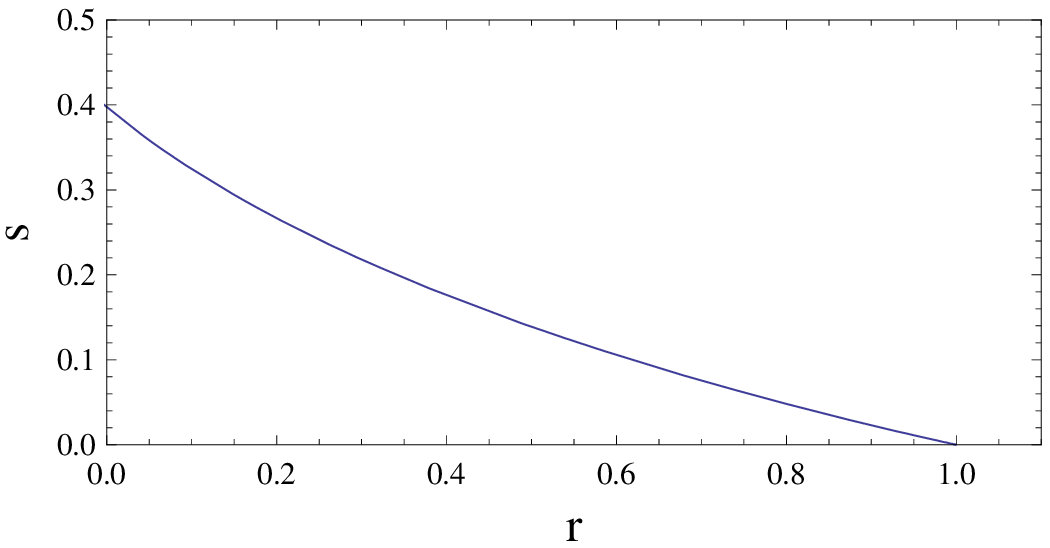}
\vspace{2mm}
~~Fig.1~~~~~~~~~~~~~~~~~~~~~~~~~~~~~~~~~~~~~~~~~~~~~~~~~~~~~~~~Fig.2~~~~~~~~~\\
\vspace{6mm} Fig. 1 shows the variations of $V$ against $\phi$,
for $A=1, B=2, k=-1,0,1,\alpha=2.5,\beta=3, m=2, n=3, d=5$ and
Fig. 2 shows the variation of the statefinder parameters $r$
against $s$ for $A=2, B=3, \alpha=3,\beta=2, m=3, n=2, d=5$ in the
case of Tachyonic field for Emergent-Powerlaw Scenario.
\vspace{6mm}
\end{figure}

\begin{figure}
\includegraphics[scale=0.4]{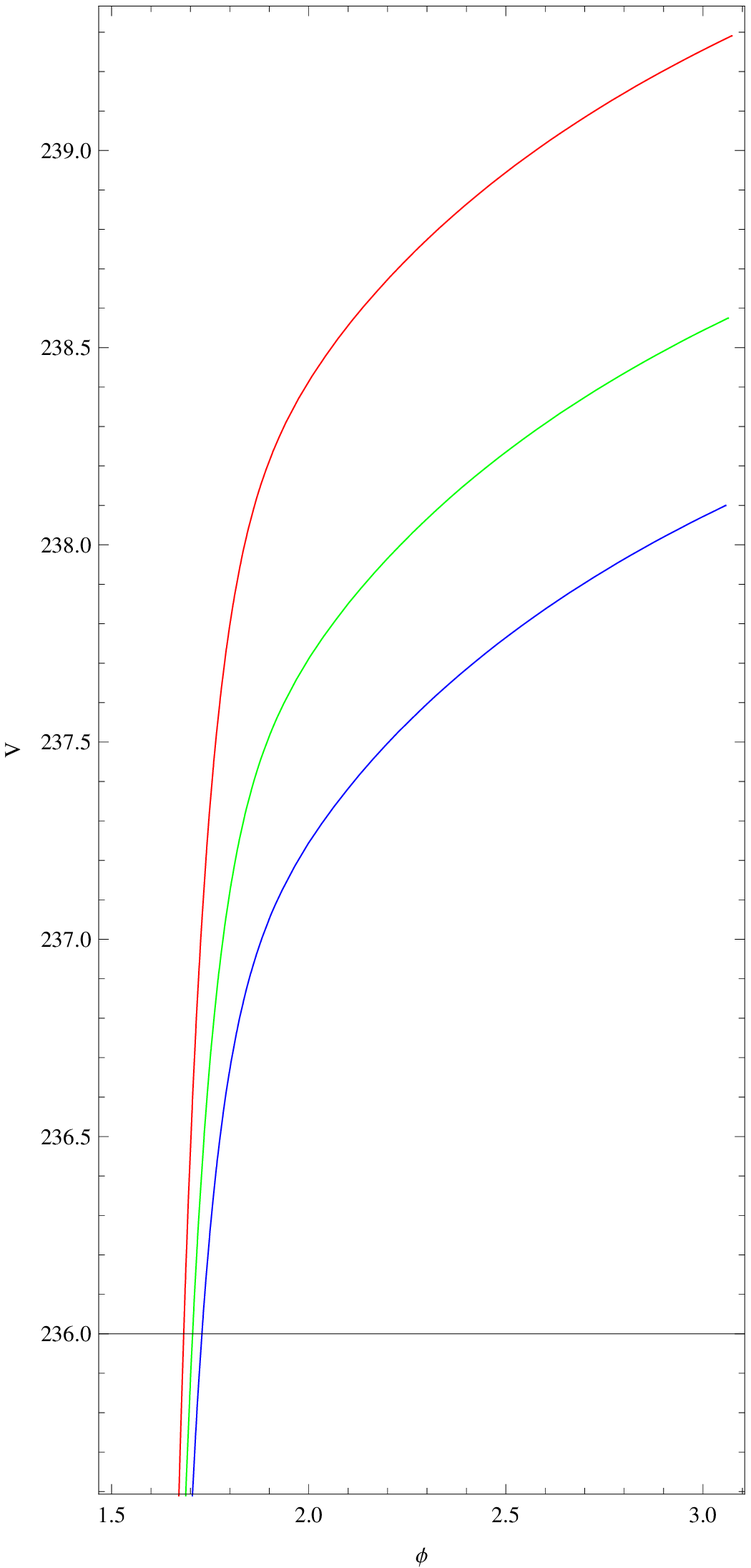}~~~~~
\includegraphics[scale=0.8]{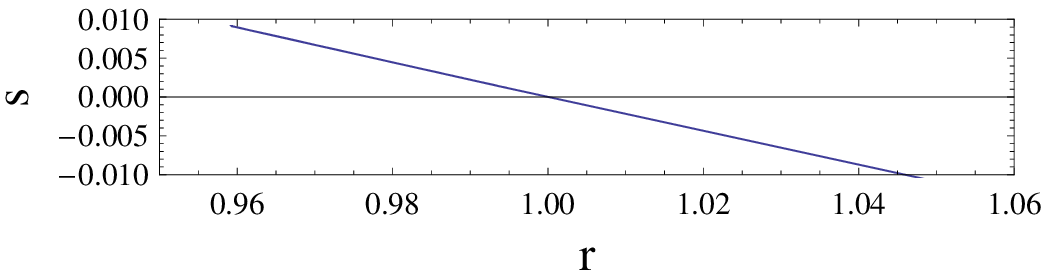}
\vspace{2mm}
~~Fig.3~~~~~~~~~~~~~~~~~~~~~~~~~~~~~~~~~~~~~~~~~~~~~~~~~~~~~~~~Fig.4~~~~~~~~~\\
\vspace{6mm} Fig. 3 shows the variations of $V$ against $\phi$,
for $A=2, B=1, k=-1,0,1,\alpha=3,\beta=2, m=3, n=.8, d=3$ and Fig.
4 shows the variation of the statefinder parameters $r$ against
$s$ for $A=2, B=3, \alpha=2,\beta=7, m=2, n=0.5, d=5$ in the case
of Tachyonic field for Emergent-Intermediate Scenario.
\vspace{6mm}
\end{figure}

\begin{figure}
\includegraphics[scale=0.4]{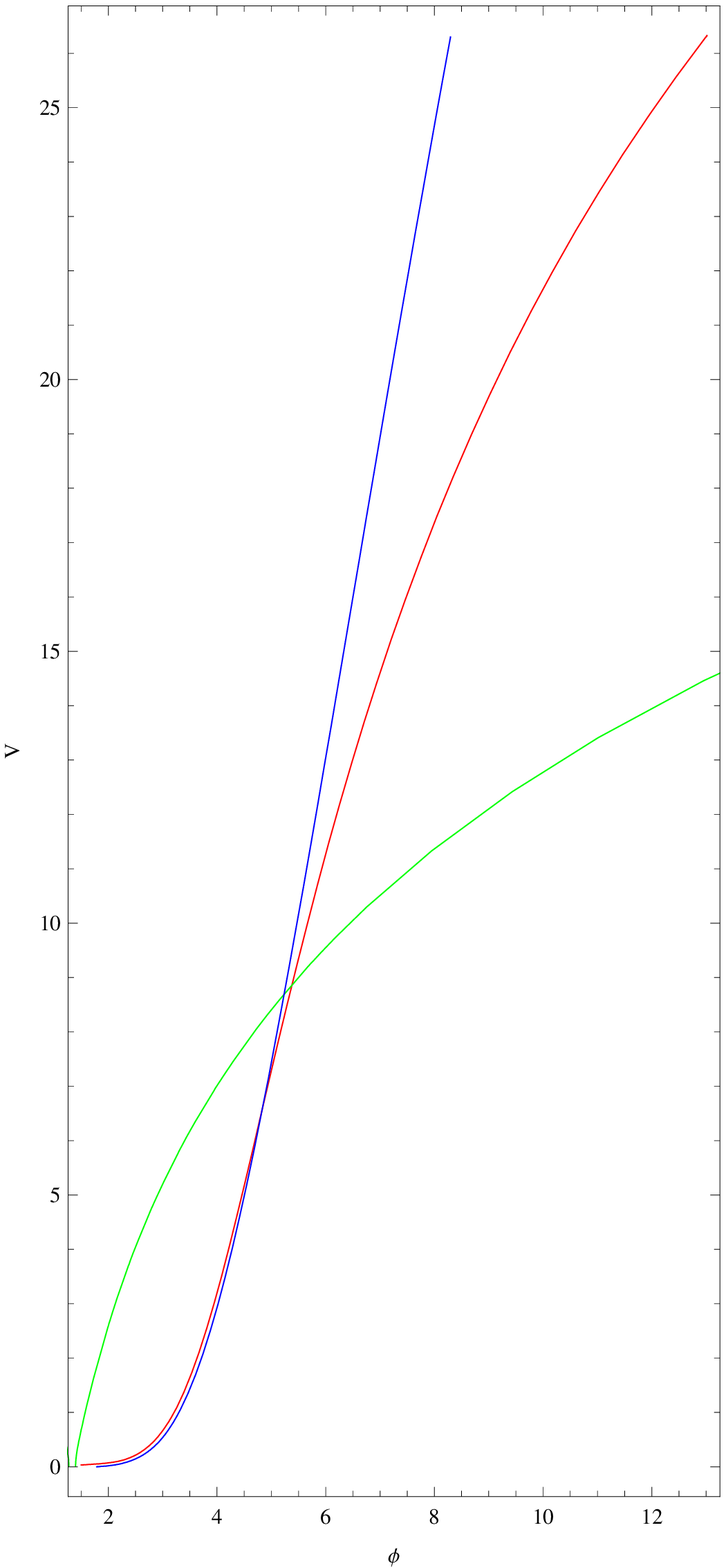}~~~~~
\includegraphics[scale=0.8]{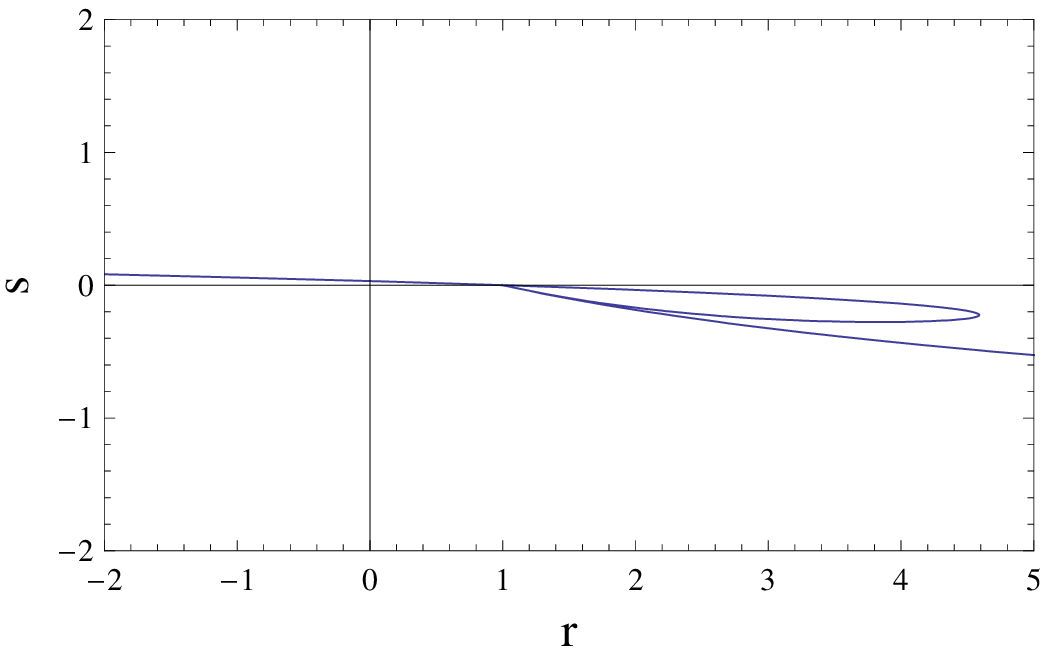}
\vspace{2mm}
~~Fig.5~~~~~~~~~~~~~~~~~~~~~~~~~~~~~~~~~~~~~~~~~~~~~~~~~~~~~~~Fig.6~~~~~~\\
\vspace{6mm} Fig. 5 shows the variations of $V$ against $\phi$,
for $A=2, B=1, k=-1,0,1,\alpha=0.1,\beta=1, m=0.3, n=5, d=3$ and
Fig. 6 shows the variation of the statefinder parameters $r$
against $s$ for $A=2, B=3, \alpha=3,\beta=2, m=3.5, n=2, d=5$ in
the case of Tachyonic field for Emergent-Logamediate Scenario.
\vspace{6mm}
\end{figure}

\begin{figure}
\includegraphics[scale=0.35]{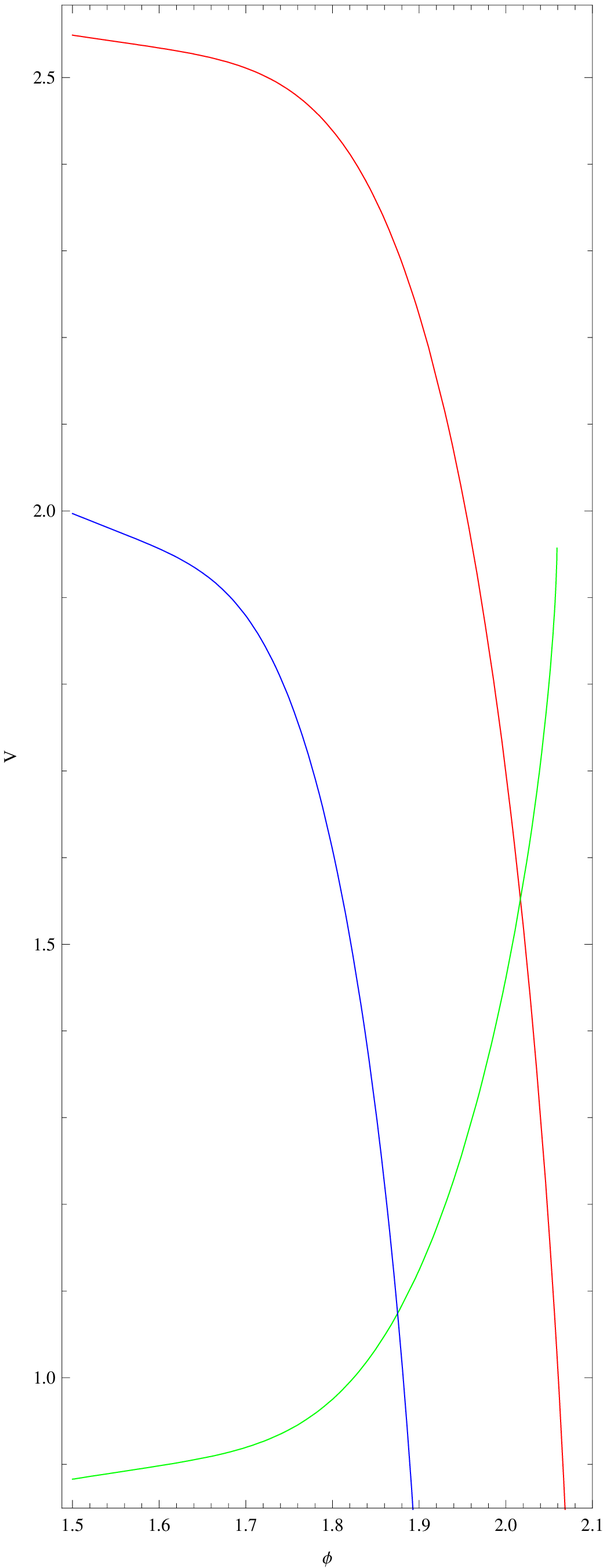}~~~~~
\includegraphics[scale=0.8]{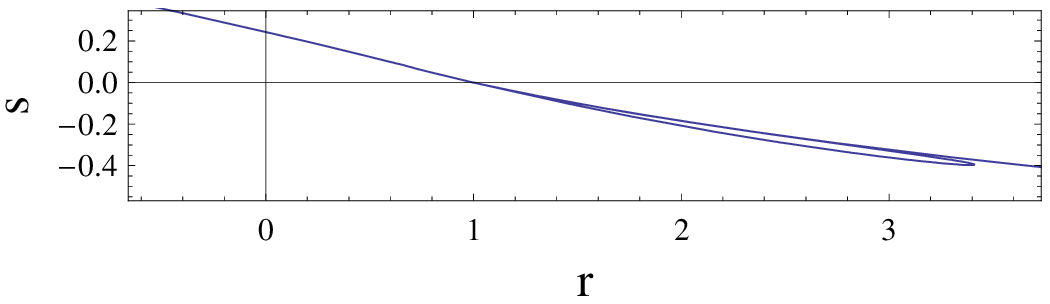}
\vspace{2mm}
~~Fig.7~~~~~~~~~~~~~~~~~~~~~~~~~~~~~~~~~~~~~~~~~~~~~~~~~~~~~~~Fig.8~~~~~~~~~~~\\
\vspace{6mm} Fig. 7 shows the variations of $V$ against $\phi$,
for $A=2, B=1, k=-1,0,1, m=10, n=0.9, d=3$ and Fig. 8 shows the
variation of the statefinder parameters $r$ against $s$ for $A=2,
B=3, m=8, n=0.8, d=5$ in the case of Tachyonic field for
Logamediate-Intermediate Scenario. \vspace{6mm}
\end{figure}

$\bullet$ \textbf{Emergent-Power law Scenario:}\\

In this section we consider a combination of scale factors $a$ and
$b$ in the emergent and power laws form respectively. As we are
considering anisotropic universe, this combination appears
feasible. We name this scenario as ``emergent-power law scenario".
In this scenario we consider $a(t)$ and $b(t)$ as follows:
\begin{equation}
 a(t)=A\left(\beta+e^{\alpha t}\right)^{m}~~\text{and} ~~
 b(t)=Bt^{n}
\end{equation}
 Using equations (4)-(6), we can find the expressions for $V(\phi)$ and
$\phi$ as
\begin{eqnarray*}
V(t)=\frac{1}{2\sqrt{2}}\sqrt{\left(-\frac{d
n(dn+n-4)}{8t^{2}}+\frac{3e^{2\alpha t}m^{2}\alpha^{2}}{(e^{\alpha
t}+\beta)^{2}}+\frac{3k(\beta+e^{\alpha t})^{-2m}}{A^{2}}\right)}
\end{eqnarray*}
\begin{equation}
\times\sqrt{\left(\frac{8k(e^{\alpha
t}+\beta)^{-2m}}{A^{2}}+\frac{8e^{\alpha t}m \alpha^{2}(3e^{\alpha
t}m+2\beta)}{(e^{\alpha
t}+\beta)^{2}}+\frac{dn(n-dn-\frac{4e^{\alpha t}m
t\alpha}{e^{\alpha t}+\beta})}{t^{2}}\right)}
\end{equation}
\begin{equation}
\phi(t)=\int{\sqrt{1+\frac{\frac{d^{2}n^{2}}{t^{2}}-\frac{8k(e^{\alpha
t}+\beta)^{-2m}}{A^{2}}-\frac{8e^{\alpha t}m \alpha^{2}(3e^{\alpha
t}m+2\beta)}{(e^{\alpha
t}+\beta)^{2}}+\frac{dn(n-dn-\frac{4e^{\alpha t}m
t\alpha}{e^{\alpha t}+\beta})}{t^{2}}}{8\left(-\frac{d
n(dn+n-4)}{8t^{2}}+\frac{3e^{2\alpha t}m^{2}\alpha^{2}}{(e^{\alpha
t}+\beta)^{2}}+\frac{3k(\beta+e^{\alpha
t})^{-2m}}{A^{2}}\right)}}}~dt
\end{equation}\\
Using equation (7), we can find the expressions for $r$ and $s$
as\\
\begin{equation}
r=1-\frac{3(d+3)(-3e^{\alpha t}m t^{2}\alpha^{2}\beta+dn(e^{\alpha
t}+\beta)^{2})}{(e^{\alpha t}(dn+3mt\alpha)+dn\beta)^{2}}+
\frac{(3+d)^{2}(3e^{\alpha t}mt^{3}\alpha^{3}\beta(-e^{\alpha
t}+\beta)+2dn(e^{\alpha t}+\beta)^{3})}{(e^{\alpha
t}(dn+3mt\alpha)+dn\beta)^{3}}
\end{equation}
\begin{eqnarray*}
s=\left[\frac{(-3(3+d)(e^{\alpha t}(d
n+3mt\alpha)+dn\beta)(-3e^{\alpha
t}mt^{2}\alpha^{2}\beta+dn(e^{\alpha
t}+\beta)^{2})+(3+d)^{2}}{(3(e^{\alpha t}(d n+3mt\alpha)+d
n\beta)^{3}}\right.
\end{eqnarray*}
\begin{equation}
\left.\frac{\times(3e^{\alpha t}mt^{3}\alpha^{3}\beta(-e^{\alpha
t}+\beta)+2dn(e^{\alpha t}+\beta)^{3}))}{\times
(-\frac{3}{2}+\frac{(3+d)(-3e^{\alpha
t}mt^{2}\alpha^{2}\beta+dn(e^{\alpha t}+\beta)^{2})}{(e^{\alpha
t}(d n+3mt\alpha)+d n\beta)^{2}}))}\right]
\end{equation}\\
The above forms of statefinder parameters pertain to the
emergent-power law scenario. The statefinder parameters derived
above are plotted as $r-s$ trajectory in the figure 2 and the
potential $V$ corresponding to this scenario is plotted against
the scalar field $\phi$ in figure 1.
\\

$\bullet$ \textbf{Emergent-Intermediate Scenario:}\\

In this section we consider a combination of scale factors $a$ and
$b$ in the forms of emergent and intermediate respectively. We dub
this scenario as ``Emergent-Intermediate Scenario". Thus, to
describe this scenario we consider $a(t)$ and $b(t)$ as follows:
\begin{equation}
 a(t)=A\left(\beta+e^{\alpha t}\right)^{m}~~\text{and} ~~
 b(t)=exp (B t^{n})
\end{equation}
Using equations (4)-(6), we can find the expressions for $V(\phi)$
and $\phi$ as\\
\begin{eqnarray*}
V(t)=\frac{1}{8}\sqrt{-\frac{e^{-2A(\ln t)^{m}}}{t^{4}(\ln
t)^{2}}((24kt^{2}-Bde^{2A(\ln
t)^{m}}nt^{n}(-4+n(4+B(d+1)t^{n})))(\ln t)^{2}+24A^{2}e^{2A(\ln
t)^{m}}m^{2}(\ln t)^{2m})}
\end{eqnarray*}
\begin{equation}
\times\sqrt{(-8e^{-2A(\ln t)^{m}}kt^{2}+B^{2}(-1+d)d
n^{2}t^{2n}+4Am(\ln t)^{m-2}(4-4m+(4+Bdn t^{n})(\ln t)-6Am(\ln
t)^{m}))}
\end{equation}

\begin{eqnarray*}
\phi(t)=\int{\sqrt{1+\frac{e^{2A(\ln t)^{m}}(\ln
t)^{2}(-8e^{-2A(\ln t)^{m}}kt^{2}+B^{2}(-1+d)d n^{2}t^{2n}+4Am(\ln
t)^{m-2}}{(24kt^{2}-Bde^{2A(\ln
t)^{m}}nt^{n}(-4+n(4+B(d+1)t^{n})))(\ln t)^{2}+24A^{2}e^{2A(\ln
t)^{m}}m^{2}(\ln t)^{2m}}}}
\end{eqnarray*}
\begin{equation}
{\overline{\frac{\times(4-4m+(4+B d n t^{n})(\ln t)-6Am(\ln
t)^{m}))}{}}}~dt
\end{equation}

Using equation (7), we can find the expressions for $r$ and $s$
as\\
\begin{eqnarray*}
r=1+\frac{3(d+3)(Bd(-1+n)nt^{n}(\ln t)^{2}+3Am(-1+m-\ln t)(\ln
t)^{m})}{(Bdnt^{n}\ln t+3Am(\ln t)^{m})^{2}}
\end{eqnarray*}
\begin{equation}
+\frac{(3+d)^{2}(Bdn(2+(-3+n)n)t^{n}(\ln t)^{3}+3Am(\ln
t)^{m}(2+(-3+m)m+\ln t(3-3m+2\ln t)))}{(Bdnt^{n}\ln t+3Am(\ln
t)^{m})^{3}}
\end{equation}
\begin{eqnarray*}
s=\left[\frac{-((3+d)(3(Bdnt^{n}\ln t+3Am(\ln
t)^{m})(Bd(-1+n)nt^{n}(\ln t)^{2}+3Am(-1+m-\ln t)(\ln
t)^{m})}{3(Bdnt^{n}\ln t+3Am(\ln
t)^{m})^{3}(\frac{3}{2}+\frac{(d+3)(Bd(-1+n)nt^{n}(\ln
t)^{2}+3Am(-1+m-\ln t)(\ln t)^{m})}{(Bdnt^{n}\ln t+3Am(\ln
t)^{m})^{2}})}\right.
\end{eqnarray*}
\begin{equation}
\left.\frac{+(3+d)(Bdn(2+(-3+n)n)t^{n}(\ln t)^{3}+3Am(\ln
t)^{m}(2+(-3+m)m+\ln t(3-3m+2\ln t)))))}{}\right]
\end{equation}\\
The forms of potential and $r-s$ trajectory derived in this
scenario are presented in figures 3 and 4 respectively.
\\

$\bullet$ \textbf{Emergent-Logamediate Scenario:}\\

In this section we consider the scale factor $a$ in emergent form
and $b$ in the logamediate form. We dub this scenario as
``Emergent-Logamediate Scenario". Thus, in this scenario we
consider $a(t)$ and $b(t)$ as follows:
\begin{equation}
 a(t)=A\left(\beta+e^{\alpha t}\right)^{m}~~\text{and} ~~
 b(t)=exp(B(\ln t)^{n})
\end{equation}
Using equations (4)-(6), we can find the expressions for $V(\phi)$
and $\phi$ as\\
\begin{eqnarray*}
V(t)=\frac{1}{8}\sqrt{(24(\frac{e^{2\alpha
t}m^{2}\alpha^{2}}{(e^{\alpha t}+\beta)^{2}}+\frac{k(e^{\alpha
t}+\beta)^{-2m}}{A^{2}})-\frac{Bdn(\ln t)^{n-2}(4(n-1)-4\ln
t)+B(1+d)n(\ln t)^{n}}{t^{2}})}
\end{eqnarray*}
\begin{equation}
\times\sqrt{(\frac{8k(e^{\alpha
t}+\beta)^{-2m}}{A^{2}}+\frac{8e^{\alpha t}m\alpha^{2}(3e^{\alpha
t}m+2\beta)}{(e^{\alpha t}+\beta)^{2}}-\frac{Bdn(\ln
t)^{n-2}(4e^{\alpha t}m t \alpha \ln t+B(-1+d)n(e^{\alpha
t}+\beta)(\ln t)^{n})}{t^{2}(e^{\alpha t}+\beta)})}
\end{equation}

\begin{equation}
\phi(t)=\int{\sqrt{\left(1+\frac{(-\frac{8k(e^{\alpha
t}+\beta)^{-2m}}{A^{2}}-\frac{8e^{\alpha t}m\alpha^{2}(3e^{\alpha
t}m+2\beta)}{(e^{\alpha t}+\beta)^{2}}+\frac{Bdn(\ln
t)^{n-2}(4e^{\alpha t}m t \alpha \ln t+B(-1+d)n(e^{\alpha
t}+\beta)(\ln t)^{n})}{t^{2}(e^{\alpha
t}+\beta)})}{(24(\frac{e^{2\alpha t}m^{2}\alpha^{2}}{(e^{\alpha
t}+\beta)^{2}}+\frac{k(e^{\alpha
t}+\beta)^{-2m}}{A^{2}})-\frac{Bdn(\ln t)^{n-2}(4(n-1)-4\ln
t)+B(1+d)n(\ln t)^{n}}{t^{2}})}\right)}}~dt
\end{equation}

Using equation (7), we can find the expressions for $r$ and $s$
as\\
\begin{eqnarray*}
r=1+\frac{3(d+3)\left(\frac{3e^{\alpha
t}m\alpha^{2}\beta}{(e^{\alpha t}+\beta)^{2}}+\frac{Bdn(\ln t
+n-1)(\ln t)^{n-2}}{t^{2}}\right)}{\left(\frac{3e^{\alpha
t}m\alpha}{e^{\alpha t}+\beta}+\frac{Bdn(\ln
t)^{n-1}}{t}\right)^{2}}
\end{eqnarray*}
\begin{equation}
+\frac{(d+3)^{2}\left(\frac{3e^{\alpha
t}m\alpha^{3}\beta(-e^{\alpha t}+\beta)}{(e^{\alpha
t}+\beta)^{3}}+\frac{Bdn(\ln t)^{n-3}(2+(-3+n)n+\ln t(3-3n+2\ln
t))}{t^{3}}\right)}{\left(\frac{3e^{\alpha t}m\alpha}{e^{\alpha
t}+\beta}+\frac{Bdn(\ln t)^{n-1}}{t}\right)^{3}}
\end{equation}

\begin{eqnarray*}
s=-\frac{(d+3)\left(\frac{3e^{\alpha
t}m\alpha^{2}\beta}{(e^{\alpha t}+\beta)^{2}}+\frac{Bdn(\ln t
+n-1)(\ln t)^{n-2}}{t^{2}}\right)}{\left(\frac{3e^{\alpha
t}m\alpha}{e^{\alpha t}+\beta}+\frac{Bdn(\ln
t)^{n-1}}{t}\right)^{2}\left(\frac{3}{2}+\frac{(d+3)\left(\frac{3e^{\alpha
t}m\alpha^{2}\beta}{(e^{\alpha t}+\beta)^{2}}+\frac{Bdn(\ln t
+n-1)(\ln t)^{n-2}}{t^{2}}\right)}{\left(\frac{3e^{\alpha
t}m\alpha}{e^{\alpha t}+\beta}+\frac{Bdn(\ln
t)^{n-1}}{t}\right)^{2}}\right)}
\end{eqnarray*}
\begin{equation}
-\frac{(d+3)^{2}\left(\frac{3e^{\alpha
t}m\alpha^{3}\beta(-e^{\alpha t}+\beta)}{(e^{\alpha
t}+\beta)^{3}}+\frac{Bdn(\ln t)^{n-3}(2+(-3+n)n+\ln t(3-3n+2\ln
t))}{t^{3}}\right)}{3\left(\frac{3e^{\alpha t}m\alpha}{e^{\alpha
t}+\beta}+\frac{Bdn(\ln
t)^{n-1}}{t}\right)^{3}\left(\frac{3}{2}+\frac{(d+3)\left(\frac{3e^{\alpha
t}m\alpha^{2}\beta}{(e^{\alpha t}+\beta)^{2}}+\frac{Bdn(\ln t
+n-1)(\ln t)^{n-2}}{t^{2}}\right)}{\left(\frac{3e^{\alpha
t}m\alpha}{e^{\alpha t}+\beta}+\frac{Bdn(\ln
t)^{n-1}}{t}\right)^{2}}\right)}
\end{equation}
In figures 5 and 6 we have plotted the tachyonic field potential
against $V$ against scalar field $\phi$ for this scenario and
$r-s$ trajectory respectively.
\\

$\bullet$ \textbf{Logamediate-Intermediate Scenario:}\\

In this section we consider $a$ in the logamediate form and $b$ in
the intermediate form. We name this scenario as
``logamediate-intermediate Scenario". Hence, in this scenario we
consider $a(t)$ and $b(t)$ as follows:
\begin{equation}
 a(t)=exp(A(\ln t)^{m})~~\text{and} ~~
 b(t)=exp (B t^{n})
\end{equation}
Using equations (4)-(6), we can find the expressions for $V(\phi)$
and $\phi$ as\\
\begin{eqnarray*}
V(t)=\frac{1}{8}
\end{eqnarray*}
\begin{eqnarray*}
\times\sqrt{\left(-\frac{e^{-2A(\ln t)^{m}}}{t^{4}(\ln
t)^{2}}((24kt^{2}-Bde^{2A(\ln
t)^{m}}nt^{n}(-4+n(4+B(1+d)t^{n})))(\ln t)^{2}+24A^{2}e^{2A(\ln
t)^{m}}m^{2}(\ln t)^{2m})\right)}
\end{eqnarray*}
\begin{equation}
\times\sqrt{\left(-8e^{-2A(\ln t)^{m}}k
t^{2}+B^{2}(-1+d)dn^{2}t^{2n}+4Am(\ln
t)^{m-2}(4-4m+(4+Bdnt^{n})\ln t-6Am(\ln t)^{m})\right)}
\end{equation}

\begin{eqnarray*}
\phi(t)=\int{\sqrt{1+\frac{e^{2A(\ln t)^{m}}(\ln
t)^{2}(-8e^{-2A(\ln t)^{m}}k t^{2}+B^{2}(-1+d)dn^{2}t^{2n}+4Am(\ln
t)^{m-2}}{(24kt^{2}-Bde^{2A(\ln
t)^{m}}nt^{n}(-4+n(4+B(1+d)t^{n})))(\ln t)^{2}+24A^{2}e^{2A(\ln
t)^{m}m^{2}(\ln t)^{2m}}}}}
\end{eqnarray*}
\begin{equation}
{\overline{\frac{\times(4-4m+(4+Bdnt^{n})\ln t-6Am(\ln
t)^{m})}{}}}~dt
\end{equation}

Using equation (7), we can find the expressions for $r$ and $s$
as\\
\begin{eqnarray*}
r=1+\frac{3(d+3)(Bd(n-1)nt^{n}(\ln t)^{2}+3Am(-ln t+m-1)(\ln
t)^{m})}{(Bdnt^{n}\ln t+3Am(\ln t)^{m})^{2}}
\end{eqnarray*}
\begin{equation}
+\frac{((d+3)^{2}(Bdn(2+(n-3)n)t^{n}(\ln t)^{3}+3Am(\ln
t)^{m}(2+(m-3)m+\ln t(3-3m+2\ln t))))}{(Bdnt^{n}\ln t+3Am(\ln
t)^{m})^{3}}
\end{equation}
\begin{eqnarray*}
s=-\frac{((d+3)(3(Bdnt^{n}\ln t+3Am(\ln t)^{m})(Bd(-1+n)nt^{n}(\ln
t)^{2}+3Am(-\ln t+m-1)(\ln t)^{m})+(d+3)}{3(Bdnt^{n}\ln t+3Am(\ln
t)^{m})^{3}(\frac{3}{2}+\frac{3(d+3)(Bd(n-1)nt^{n}(\ln
t)^{2}+3Am(-ln t+m-1)(\ln t)^{m})}{(Bdnt^{n}\ln t+3Am(\ln
t)^{m})^{2}})}
\end{eqnarray*}
\begin{equation}
\frac{\times(Bdn(2+(-3+n)n)t^{n}(\ln t)^{3}+3Am(\ln
t)^{m}(2+(-3+m)m+\ln t(3-3m+2\ln t)))))}{}
\end{equation}

In figures 7 and 8 we have plotted the tachyonic field potential
against $V$ against scalar field $\phi$ for this
``logamediate-intermediate scenario" and $r-s$ trajectory
respectively.
\\
\section{\normalsize\bf{Discussions}}

In this work, we have considered $N~(=4+d)$-dimensional Einstein's
field equations in which 4-dimensional space-time is described by
a FRW metric and that of the extra $d$-dimensions by an Euclidean
metric. We have considered four scenarios, namely,
\emph{Emergent-Powerlaw,
Emergent-Intermediate,Emergent-Logamediate and
Logamediate-Intermediate} scenarios where the universe is filled
with Tachyonic Field dark energy. The natures of the potentials as
well as dynamics of scalar fields for the Tachyonic Field dark
energy models have been analyzed. The statefinder parameters have
been considered and their natures have been investigated for four
reconstructed scenarios of the universe.\\

In the case of \emph{Emergent-Power law} scenario, we have
reconstructed particular forms of scale factors $a$ and $b$ where
$a$ belongs to Emergent Scenario and $b$ belongs to Powerlaw
scenario in such a way that there is no singularity for evolution
of the anisotropic Universe. We have found $\phi$ and potential
$V$ in terms of cosmic time $t$ for Tachyonic Field. Here we have
shown that the \emph{Emergent-Powerlaw} scenario is possible for
open, closed or flat Universe if the Universe contains Tachyonic
Field. From figure 1 it has been seen that the potential always
increases with
 Tachyonic Field. The $\{r,s\}$ diagram (fig.2) shows that the
  evolution of the \emph{Emergent-Powerlaw} Universe starts
from asymptotic Einstein's static era ($r\rightarrow
\infty,~s\rightarrow -\infty$) and goes to $\Lambda$CDM model
($r=1,~s=0$). It is also observed that $r,s$ are independent of
the dimension $d$. So, from statefinder parameters, the behavior
of different stages of the evolution of the Emergent-Powerlaw Universe
 have been generated.\\

In the case of \emph{Emergent-Intermediate} scenario, we have
reconstructed particular forms of scale factors $a$ and $b$ where
$a$ belongs to Emergent Scenario and $b$ belongs to Intermediate
scenario in such a way that there is no singularity for evolution
of the anisotropic Universe. We have found $\phi$ and potential
$V$ in terms of cosmic time $t$ for Tachyonic Field. Here we have
shown that the Emergent-Intermediate scenario is possible for
open, closed or flat Universe if the Universe contains Tachyonic
Field. From figure 3 it has been seen that the potential always
increases with Tachyonic Field. The $\{r,s\}$ diagram (fig.4)
shows that the evolution of the Emergent-Intermediate Universe
starts from asymptotic Einstein's static era ($r\rightarrow
\infty,~s\rightarrow -\infty$) and goes to $\Lambda$CDM model
($r=1,~s=0$). It is also observed that $r,s$ are independent of
the dimension $d$. So, from statefinder parameters, the behavior
of different stages of the evolution of
the Emergent-Intermediate Universe have been generated.\\

In the case of \emph{Emergent-Logamediate} scenario, we have
reconstructed particular forms of scale factors $a$ and $b$ where
$a$ belongs to Emergent Scenario and $b$ belongs to Logamediate
scenario in such a way that there is no singularity for evolution
of the anisotropic Universe. We have found $\phi$ and potential
$V$ in terms of cosmic time $t$ for Tachyonic Field. Here we have
shown that the Emergent-Intermediate scenario is possible for
open, closed or flat Universe if the Universe contains Tachyonic
Field. From figure 5 it has been seen that the potential always
increases with Tachyonic Field. The $\{r,s\}$ diagram (fig.6)
shows that the evolution of the Emergent-Logamediate Universe
starts from asymptotic Einstein's static era ($r\rightarrow
\infty,~s\rightarrow -\infty$) and goes to $\Lambda$CDM model
($r=1,~s=0$). It is also observed that $r,s$ are independent of
the dimension $d$. So, from statefinder parameters, the behavior
of different stages of the evolution of
the Emergent-Logamediate Universe have been generated.\\

In the case of \emph{Logamediate-Intermediate} scenario, we have
reconstructed a particular forms of scale factors $a$ and $b$
where $a$ belongs to Logamediate Scenario and $b$ belongs to
Intermediate scenario in such a way that there is no singularity
for evolution of the anisotropic Universe. We have found $\phi$
and potential $V$ in terms of cosmic time $t$ for Tachyonic Field.
Here we have shown that the Logamediate-Intermediate scenario is
possible for open, closed or flat Universe if the Universe
contains Tachyonic Field. From figure 7 it has been seen that the
potential always decreases with Tachyonic Field. The $\{r,s\}$
diagram (fig.8) shows that the evolution of the
Logamediate-Intermediate Universe starts from asymptotic
Einstein's static era ($r\rightarrow \infty,~s\rightarrow
-\infty$) and goes to $\Lambda$CDM model ($r=1,~s=0$). It is also
observed that $r,s$ are independent of the dimension $d$. So, from
statefinder parameters, the behavior of different stages of the
evolution of the Logamediate-Intermediate Universe have been generated.\\

\section{\normalsize\bf{Acknowledgments}}
The authors are thankful to the reviewers for providing
constructive and thoughtful suggestions to enhance the quality of
the paper. The second author is thankful to the Inter-University
Centre for Astronomy and Astrophysics (IUCAA), Pune for providing
Visiting Associateship to him.
\\

\end{document}